\documentclass[aps,showpacs,nofootinbib,twocolumn]{revtex4}

\usepackage{latexsym}
\usepackage{epsfig}
\usepackage{amssymb}

\newcommand{\ba}{\begin{eqnarray}}
\newcommand{\ea}{\end{eqnarray}}
\newcommand{\be}{\begin{equation}}
\newcommand{\ee}{\end{equation}}

\newcommand{\ka}{\kappa}

\newcommand{\R}{\mathcal{R}}

\begin{document}

\title{Wormholes supported by hybrid metric-Palatini gravity}

\author{Salvatore Capozziello$^{1,2}$}\email{capozzie@na.infn.it}
\author{Tiberiu Harko$^3$}\email{harko@hkucc.hku.hk}
\author{Tomi S. Koivisto$^{4}$}\email{tomi.koivisto@fys.uio.no}
\author{Francisco S.N.~Lobo$^{5}$}\email{flobo@cii.fc.ul.pt}
\author{Gonzalo J. Olmo$^{6}$}\email{gonzalo.olmo@csic.es}

\affiliation{$^1$Dipartimento di Scienze Fisiche, Universit\`{a} di Napoli "Federico II",
 Compl. Univ. di Monte S. Angelo, Edificio G, Via Cinthia, I-80126, Napoli, Italy}
\affiliation{$^2$ INFN Sez. di Napoli, Compl. Univ. di Monte S. Angelo, Edificio G, Via Cinthia, I-80126, Napoli, Italy}
\affiliation{$^3$Department of Physics and Center for Theoretical
and Computational Physics, The University of Hong Kong, Pok Fu Lam Road, Hong Kong}
\affiliation{$^{4}$  Institute for Theoretical Astrophysics, University of
  Oslo, P.O.\ Box 1029 Blindern, N-0315 Oslo, Norway}
\affiliation{$^5$Centro de Astronomia e Astrof\'{\i}sica da Universidade de Lisboa, Campo Grande, Ed. C8 1749-016 Lisboa, Portugal}
\affiliation{$^6$Departamento de F\'{i}sica Te\'{o}rica and IFIC, Centro Mixto Universidad de
Valencia - CSIC. Universidad de Valencia, Burjassot-46100, Valencia, Spain}

\date{\today}

\begin{abstract}

Recently, a modified theory of gravity was presented, which consists of the superposition of the metric Einstein-Hilbert Lagrangian with an $f(\R)$ term constructed \`{a} la Palatini. The theory possesses extremely interesting features such as predicting the existence of a long-range scalar field, that explains the late-time cosmic acceleration and passes the local tests, even in the presence of a light scalar field. In this brief report, we consider the possibility that wormholes are supported by this hybrid metric-Palatini gravitational theory. We present here the general conditions for  wormhole solutions according to  the null energy conditions at the throat and find specific examples. In the first solution, we specify the redshift function, the scalar field and choose the potential that simplifies the modified Klein-Gordon equation. This solution is not asymptotically flat and needs to be matched to a vacuum solution. In the second example, by adequately specifying the metric functions and choosing the scalar field, we find an asymptotically flat spacetime.

\end{abstract}


\maketitle

\section{Introduction}

Wormholes are hypothetical shortcuts in spacetime and are primarily useful as ``gedanken experiments'' and as a  theoretical probe on  the foundation of general relativity \cite{Morris:1988cz}. In classical general relativity, it is well-known that wormholes possess a peculiar property, namely `exotic matter,' involving a stress-energy tensor $T_{\mu\nu}$ that violates the null energy condition (NEC) at the throat, i.e., $T_{\mu\nu}k^\mu k^\nu < 0$, where $k^\mu$ is any null vector. However, it has been recently shown that in the context of modified gravity, the stress-energy tensor of standard matter can be imposed to satisfy the usual energy conditions and it is the higher order curvature terms  that support these exotic geometries \cite{wh_modgrav}.

In fact, modified gravity \cite{fRgravity} has mainly been revived to explain the late-time cosmic acceleration. Indeed, generalizations of these modified gravitational theories have also been extensively analyzed in the literature, such as C-theories
\cite{Amendola:2010bk}, nonminimal curvature-matter couplings \cite{Harko:2010mv}, etc. A natural way to obtain solely gravitational modifications of the behavior of matter emerges in the Palatini
formulation of extended gravity actions \cite{Olmoetal}. In the latter, the relation between the independent connection and the metric depends upon the trace of the matter stress-energy tensor in such a way that the field equations effectively feature extra terms given by the matter content.
However, since the extra terms are fourth order in (spatial) derivatives \cite{Olmo:2011fh}, some models of these theories are problematical both at the theoretical and phenomenological levels \cite{Koivisto:2005yc}.

In this context, a novel approach to modified theories of gravity was recently proposed, consisting of adding to the metric Einstein-Hilbert Lagrangian an $f(\R)$ term constructed \`{a} la Palatini \cite{Harko:2011nh}. It was shown that using the respective dynamically equivalent scalar-tensor representation,
the theory passes the Solar System observational constraints even
if the scalar field is very light and also leads to the late-time cosmic acceleration. Cosmological applications have also been extensively analyzed \cite{Capozziello:2012ny}.

In this brief report, we consider the possibility that wormhole geometries are supported by the above-mentioned hybrid metric-Palatini theory. We present the generic conditions relative to the NEC and provide two specific solutions.

\section{Hybrid metric-Palatini gravity}\label{sect2}

The action for the hybrid metric-Palatini gravity  is
\begin{equation} \label{eq:S_hybrid}
S=\frac{1}{2\kappa^2}\int d^4 x \sqrt{-g} \left[ R + f(\R)\right] +
S_m,
\end{equation}
where $\kappa^2\equiv 8\pi G$, $S_m$ is the matter action, $R$ is
the metric Einstein-Hilbert term, $\R  \equiv g^{\mu\nu}\R_{\mu\nu} $ is
the Palatini curvature, and $\R_{\mu\nu}$ is defined in terms of
an independent connection $\hat{\Gamma}^\alpha_{\mu\nu}$  as
\begin{equation}
\R_{\mu\nu} \equiv \hat{\Gamma}^\alpha_{\mu\nu ,\alpha} -
\hat{\Gamma}^\alpha_{\mu\alpha , \nu} +
\hat{\Gamma}^\alpha_{\alpha\lambda}\hat{\Gamma}^\lambda_{\mu\nu}
-\hat{\Gamma}^\alpha_{\mu\lambda}\hat{\Gamma}^\lambda_{\alpha\nu}\,.
\end{equation}

Varying the action (\ref{eq:S_hybrid}) with respect to the metric,
one obtains the following gravitational field equations  \be
\label{efe} G_{\mu\nu} +
F(\R)\R_{\mu\nu}-\frac{1}{2}f(\R)g_{\mu\nu} = \ka^2 T_{\mu\nu}\,,
\ee
where the matter stress-energy tensor is defined as $ T_{\mu\nu}
\equiv -\left(2/\sqrt{-g}\right) \delta
(\sqrt{-g}\mathcal{L}_m)/\delta g^{\mu\nu}$. The
independent connection is compatible with the metric
$F(\R)g_{\mu\nu}$, conformal to $g_{\mu\nu}$; the conformal
factor is given by $F(\R) \equiv df(\R)/d\R$. The latter considerations imply that
\ba
\label{ricci} \R_{\mu\nu} & = & R_{\mu\nu} +
\frac{3}{2}\frac{1}{F^2(\R)}F(\R)_{,\mu}F(\R)_{,\nu}
    \nonumber \\
 && - \frac{1}{F(\R)}\nabla_\mu F(\R)_{,\nu} -
\frac{1}{2}\frac{1}{F(\R)}g_{\mu\nu}\Box F(\R)\,. \ea
Note that $\R$ can be obtained from the trace of the field equations (\ref{efe}), which yields $F(\R)\R -2f(\R) - R = \ka^2 T$.


Introducing an auxiliary field, the hybrid metric-Palatini action (\ref{eq:S_hybrid}) can be turned into a scalar-tensor theory given by the following action (we refer the reader to \cite{Harko:2011nh} for more details)
\begin{equation} \label{eq:S_scalar1}
S= \frac{1}{2\kappa^2}\int d^4 x \sqrt{-g} \left[ R + \phi\R-V(\phi)\right] +S_m \ .
\end{equation}
Varying this action with respect to the metric, the scalar $\phi$ and the connection yields the following field equations
\begin{eqnarray}
R_{\mu\nu}+\phi \R_{\mu\nu}-\frac{1}{2}\left(R+\phi\R-V\right)g_{\mu\nu}&=&\kappa^2 T_{\mu\nu} \,,
\label{eq:var-gab}\\
\R-V_\phi&=&0 \,, \label{eq:var-phi}\\
\hat{\nabla}_\alpha\left(\sqrt{-g}\phi g^{\mu\nu}\right)&=&0 \,, \label{eq:connection}\
\end{eqnarray}
respectively. Note that the solution of Eq.~(\ref{eq:connection}) implies that the independent connection is the Levi-Civita connection of
a metric $h_{\mu\nu}=\phi g_{\mu\nu}$. Thus we are dealing with a bi-metric theory and $\R_{\mu\nu}$ and $R_{\mu\nu}$ are related by
\begin{equation} \label{eq:conformal_Rmn}
\R_{\mu\nu}=R_{\mu\nu}+\frac{3}{2\phi^2}\partial_\mu \phi \partial_\nu \phi-\frac{1}{\phi}\left(\nabla_\mu
\nabla_\nu \phi+\frac{1}{2}g_{\mu\nu}\Box\phi\right) \ ,
\end{equation}
and consequently
$\R=R+\frac{3}{2\phi^2}\partial_\mu \phi \partial^\mu \phi-\frac{3}{\phi}\Box \phi$,
which can be used in the action (\ref{eq:S_scalar1}) to get rid of the independent connection and obtain the following
scalar-tensor representation \cite{Koivisto:2009jn}
\begin{eqnarray} \label{eq:S_scalar2}
S= \frac{1}{2\kappa^2}\int d^4 x \sqrt{-g} \left[ (1+\phi)R +\frac{3}{2\phi}\partial_\mu \phi \partial^\mu \phi
-V(\phi)\right] +S_m  .
\nonumber
\end{eqnarray}
It is important to note that this action differs fundamentally from the $w=-3/2$ Brans-Dicke theory in the coupling of the scalar to the curvature.

Now substituting Eq.~(\ref{eq:var-phi}) and Eq.~(\ref{eq:conformal_Rmn}) in Eq.~(\ref{eq:var-gab}), the metric field equation can be
written as an effective Einstein field equation, i.e.,  $G_{\mu\nu}=\kappa^2 T^{\rm eff}_{\mu\nu}$, where the effective stress-energy tensor is given by
\begin{eqnarray}
T^{\rm eff}_{\mu\nu}&=&\frac{1}{1+\phi} \Big\{ T_{\mu\nu}
 - \frac{1}{\kappa^2} \Big[ \frac{1}{2}g_{\mu\nu}\left(V+2\Box\phi\right)+
     \nonumber \\
&& \nabla_\mu\nabla_\nu\phi-\frac{3}{2\phi}\partial_\mu \phi
\;\partial_\nu \phi + \frac{3}{4\phi}g_{\mu\nu}(\partial \phi)^2 \Big] \  \Big\}  \label{effSET} .
\end{eqnarray}

The scalar field is governed by the second-order evolution equation (we refer the reader to \cite{Harko:2011nh} for more details)
\begin{equation}\label{eq:evol-phi}
-\Box\phi+\frac{1}{2\phi}\partial_\mu \phi \partial^\mu
\phi+\frac{\phi[2V-(1+\phi)V_\phi]} {3}=\frac{\phi\kappa^2}{3}T\,,
\end{equation}
which is an effective Klein-Gordon equation.
This last expression shows that, unlike in the Palatini ($w=-3/2$)
case, the scalar field is dynamical. Thus, the theory is not
affected by the microscopic instabilities that arise in Palatini
models with infrared corrections \cite{Olmo:2011uz}.

\section{Wormholes in hybrid metric-Palatini gravity}

Consider the following line element in curvature coordinates, which represents a traversable wormhole geometry \cite{Morris:1988cz}
\begin{eqnarray}
ds^2=-e^{\Phi(r)}dt^2 + \frac{dr^2}{1-b(r)/r} + r^2 \left(d \theta^2 + \sin ^2\theta  d\varphi^2  \right)
   \label{whmetric}
\end{eqnarray}
where the metric functions $b(r)$ and $\Phi(r)$ are functions of the radial coordinate, and  denoted the shape and the redshift functions, respectively. 
The radial coordinate $r$ ranges from a minimum value $r_0$, the wormhole throat, to infinity.
In order to avoid the presence of event horizons, one imposes  that $\Phi(r)$ is finite $\forall r$. It is possible to construct asymptotically flat spacetimes, in which $b(r)/r \rightarrow 0$ and $\phi \rightarrow 0$ as $r \rightarrow \infty$. Now, a fundamental ingredient in wormhole physics is the flaring-out condition of the wormhole throat $b(r_0)=r_0$ \cite{Morris:1988cz}, given by the condition $(b' r-b)/2b^2 < 0$. In general relativity, the latter condition implies that through the Einstein field equation, the stress-energy tensor violates the NEC at the throat, i.e., $T_{\mu\nu}k^\mu k^\nu |_{r_0} < 0$.

In modified gravity, it is the effective stress-energy tensor that violates the NEC at the throat, $T^{\rm eff}_{\mu\nu} k^\mu k^\nu |_{r_0} < 0$. The latter provides the following constraint in the present hybrid metric-Palatini gravitational theory
\begin{eqnarray}
T^{\rm eff}_{\mu\nu}k^\mu k^\nu |_{r_0} &=&\frac{1}{1+\phi} \Big\{ T_{\mu\nu} k^\mu k^\nu
 - \frac{1}{\kappa^2} \Big[k^\mu k^\nu \nabla_\mu\nabla_\nu\phi
     \nonumber \\
&& -\frac{3}{2\phi}k^\mu k^\nu\,\partial_\mu \phi
\;\partial_\nu \phi \Big] \  \Big\} \Big |_{r_0} < 0 \label{NECeffSET} .
\end{eqnarray}

Assuming that $1+\phi> 0$ and that standard matter satisfies the energy conditions and, in particular, the NEC, i.e, $T_{\mu\nu} k^\mu k^\nu > 0$,  one finds the generic constraint for hybrid metric-Palatini wormhole geometries
\begin{eqnarray}
0< T_{\mu\nu} k^\mu k^\nu |_{r_0} <  \frac{1}{\kappa^2} \Big[k^\mu k^\nu \nabla_\mu\nabla_\nu\phi
 - \frac{3}{2\phi}k^\mu k^\nu\,\partial_\mu \phi
\;\partial_\nu \phi \Big]  \Big|_{r_0}.  \label{NECeffSET2}
\end{eqnarray}

Using the metric (\ref{whmetric}), the effective Einstein field equation provides the following gravitational field equations
\begin{eqnarray}
\kappa^2 \rho(r) &=& \frac{b'}{r^2}(1+\phi) - \left(1-\frac{b}{r} \right)
\left[\phi'' -\frac{3(\phi')^2}{4\phi}  \right]
    \nonumber  \\
&&+\frac{\phi'}{2r}\left( b' + \frac{3b}{r}-4  \right) -\frac{V}{2} \,,
  \label{hybrid_rho}
  \\
\kappa^2 p_r(r)&=& \left[-\frac{b}{r^3}+ \frac{2\Phi'}{r}\left(1-\frac{b}{r} \right)\right](1+\phi)
   \nonumber  \\
&& + \phi' \left(\Phi' + \frac{2}{r}
 + \frac{3 \phi'}{4\phi} \right)
\left( 1-\frac{b}{r} \right) + \frac{V}{2} \,, \\
\kappa^2 p_t(r)&=&\Bigg[ \left( \Phi'' + (\Phi')^2 + \frac{\Phi'}{r}  \right)
\left(1-\frac{b}{r} \right)
   \nonumber   \\
&&\hspace{-1.7cm} + \frac{b-b'r}{2r^3}\left(1+r\Phi' \right) \Bigg] (1+\phi)
+ \left[\phi''+\phi'\Phi' + \frac{3(\phi')^2}{4\phi} \right]\times
  \nonumber   \\
  &&
\hspace{-1cm}\times
 \left( 1- \frac{b}{r} \right)
   +\frac{\phi'}{r}\left(1- \frac{b+rb'}{2r}  \right)  + \frac{V}{2} \,.
\end{eqnarray}

The effective Klein-Gordon equation (\ref{eq:evol-phi}) is given by
\begin{eqnarray}
\left[ \phi''+ \phi' \Phi' -\frac{(\phi')^2}{2\phi} + \frac{3\phi'}{2r}  \right] \left(1-\frac{b}{r} \right) + \frac{\phi'}{2r}(1+b')
   \nonumber \\
  \frac{\phi}{3}\left[ 2V - (1+\phi)V_\phi \right]
  = \frac{\phi \kappa^2}{3} T \,.
        \label{modKGeq}
\end{eqnarray}
Note that Eqs.~(\ref{hybrid_rho})-(\ref{modKGeq}) provide four independent equations, for seven unknown quantities, i.e. $\rho(r)$, $p_r(r)$, $p_t(r)$, $\Phi(r)$, $b(r)$, $\phi(r)$ and $V(r)$. Thus, the system of equations is under-determined, so that we will reduce the number of unknown functions by assuming suitable conditions. We will consider specific solutions in the next section.

\section{Specific examples}
\subsection{Solution I}

Consider for simplicity a zero redshift function and a specific choice for the scalar field:
\begin{eqnarray}
\Phi(r)=0\,, \qquad \phi(r)=\phi_0 \left(\frac{r_0}{r} \right)^\alpha
  \,.
\end{eqnarray}
In order to simplify the modified Klein-Gordon equation, consider $2V - (1+\phi)V_\phi=0$, which yields the potential
\begin{eqnarray}
V(\phi)=V_0(1+\phi)^2\,.
\end{eqnarray}
Now, substituting these choices into the expressions for the stress-energy tensor components, one obtains the following expression of the stress-energy tensor trace
\begin{eqnarray}
T&=& \frac{1}{2\kappa^2 r^2} \Bigg\{3\alpha \, \phi_0 \left(\frac{r_0}{r} \right)^\alpha \left[ 1-b'+\left( 1-\frac{b}{r} \right)(1-\alpha) \right]
    \nonumber   \\
&& \hspace{-0.75cm} + \left[ V_0 r^2 \left( 1+\phi_0 \left(\frac{r_0}{r} \right)^\alpha
\right)-b'  \right] \left[ 1+ \phi_0 \left(\frac{r_0}{r} \right)^\alpha
\right]  \Bigg\}   \,.
\end{eqnarray}
Finally, substituting these expressions into the modified Klein Gordon equation, i.e., Eq.~(\ref{modKGeq}), the latter simplifies to the following ordinary differential equation
\begin{eqnarray}
(4-3\alpha )b' + 3\alpha \left[ 1+(1-\alpha )\left( 1-\frac{b}{r} \right) \right]
   \nonumber  \\
-4V_0 r^2 \left[ 1+\phi_0 \left(\frac{r_0}{r} \right)^\alpha \right]=0\,,
\end{eqnarray}
which yields the following solution for the shape function
\begin{eqnarray}
b(r)&=&\Bigg[ \frac{2V_0 \phi_0 r^3}{3\alpha^2-8\alpha +6}\left(\frac{r_0}{r}\right)^\alpha +\frac{3\alpha(\alpha-2)r}{3\alpha^2-6\alpha+4}
   \nonumber \\
&&+\frac{4V_0r^3}{3(\alpha^2-4\alpha+4)} +r^{\frac{3\alpha(\alpha-1)}{3\alpha-4}}\, C  \Bigg]  \,,
\end{eqnarray}
where $C$ is a constant of integration, fixed by the boundary condition $b(r_0)=r_0$, which provides
\begin{eqnarray}
C&=&r_0 r_0^{\frac{-3\alpha(\alpha-1)}{3\alpha-4}}\,
\Bigg\{1- \Bigg[ \frac{2V_0 \phi_0 r_0^2}{3\alpha^2-8\alpha +6}
   \nonumber \\
&&
+\frac{3\alpha(\alpha-2)}{3\alpha^2-6\alpha+4}
   +\frac{4V_0r_0^2}{3(\alpha^2-4\alpha+4)}  \Bigg] \Bigg\}\,.
\end{eqnarray}
For the case of $\alpha =1$, the shape function is given by
\begin{eqnarray}
b(r)&=&\frac{4V_0}{3}\left( r^3 - r_0^3 \right) + \nonumber\\
&&2r_0V_0\phi_0 \left( r^2 - r_0^2 \right) + 4r_0 - 3r .
\end{eqnarray}
For this case the stress-energy tensor profile is given by
\begin{eqnarray}
\rho(r)&=& \frac{1}{6\kappa^2 r^4}\Big\{ 21V_0 r^4 +28 \phi_0 V_0 r^3 +6r^2(V_0 r_0^2 \phi_0^2-3)
   \nonumber   \\
&&   +\phi_0 r_0^2 \left[ V_0 r_0^2 (2+3\phi_0) -6   \right] \Big\}\,,
  \\
p_r(r)&=& - \frac{1}{6\kappa^2 r^4}\Big\{ 5 V_0 r^4 + 4 \phi_0 V_0 r^3 - 6r^2(V_0 r_0^2 \phi_0^2 +3)
   \nonumber   \\
&& -4r_0 r (2V_0 r_0^2 + 3 \phi_0 V_0 r_0^2-6 -3\phi_0 )
   \nonumber   \\
&& + \phi_0 r_0^2 \left[ V_0 r_0^2 (2+3\phi_0) -6   \right] \Big\} \,,
    \\
p_t(r)&=& - \frac{1}{6\kappa^2 r^4}\Big\{ 5 V_0 r^4 + 2 \phi_0 V_0 r^3
   \nonumber   \\
&& -2r_0 (2V_0 r_0^2 + 3 \phi_0 V_0 r_0^2-6 -3\phi_0 )
   \nonumber   \\
&& - \phi_0 r_0^2 \left[ V_0 r_0^2 (2+3\phi_0) -6   \right] \Big\} \,.
\end{eqnarray}
Note that this solution is not asymptotically flat, so that, in principle, we need to match the interior wormhole solution to an exterior vacuum spacetime, at a junction interface, much in the spirit of \cite{matching}.  To avoid the presence of exotic matter on the thin shell,  one may also impose, in principle, that the surface stresses, lying on the junction interface, satisfy the energy conditions.

\subsection{Solution II: Asymptotically flat spacetime}

Consider the choices for the metric functions and scalar field
\begin{eqnarray}
\Phi(r)=\Phi_0 \left(\frac{r_0}{r} \right)^\alpha \,,\;  b(r)=r_0 \left(\frac{r_0}{r} \right)^\beta \,,\; \phi(r)=\phi_0 \left(\frac{r_0}{r} \right)^\gamma \,,
 \label{def_phi}
\end{eqnarray}
respectively, where $\alpha > 0$, $\beta > -1$ and $\gamma > 0$ are constant parameters. 
We emphasize that general solutions for these specific cases can be found, however, they are extremely lengthy, so that without a significant loss of generality, we consider $\alpha=0$, $\beta=0$ and $\gamma=3$.

As before, let us insert these functions into the field equations, deduce the trace of  stress-energy tensor, given by 
%
$T=\frac{1}{2\kappa^2 r^2}\left\{4V(r)+9\phi_0 \left(\frac{r_0}{r} \right)^3
\left[ 1-2\frac{r_0}{r}  \right] \right\}\,,$
%
and substitute the latter into the modified Klein Gordon equation, which simplifies to
\begin{equation}
2r_0^3 V'(r)+27\phi_0 \left(\frac{r}{r_0}\right)^6\left [1-2\left(\frac{r_0}{r} \right) \right]=0 \,.
\end{equation}
This ordinary differential equation yields the following solution for the potential in parametric form, $V(r)$
\begin{eqnarray}
V(r)=\frac{9\phi_0}{10r_0^2}\left(\frac{r_0}{r}\right)^5\left[5\left( \frac{r_0}{r} \right) - 3  \right] \,.
\end{eqnarray}
Now, from the definition of the scalar field, Eq. (\ref{def_phi}), and with $\gamma=3$, one obtains that $r_0/r=(\phi/\phi_0)^{1/3}$, so that the potential $V(\phi)$ is finally given by
\begin{equation}
V(\phi)= \frac{9\phi_0}{2 r_0^2}\left(\frac{\phi}{\phi_0} \right)^{5/3}
\left[\left(\frac{\phi}{\phi_0} \right)^{1/3} - \frac{3}{5} \right] \,.
\end{equation}

The stress-energy tensor components have the following profile
\begin{eqnarray}
\rho(r)&=& \frac{21\phi_0}{10\kappa^2 r_0^2}\left( \frac{r_0}{r} \right)^5 \left[1 - \frac{5}{7} \left(\frac{r_0}{r}\right) \right] \,,
  \nonumber \\
p_r(r)&=& - \frac{1}{\kappa^2 r_0^2} \left( \frac{r_0}{r} \right)^3 \left\{1 +  \frac{3\phi_0}{5}\left(\frac{r_0}{r}\right)^2 \left[1- \frac{5}{6} \left(\frac{r_0}{r}\right) \right] \right\} \,,
 \nonumber  \\
p_t(r)&=& \frac{1}{2\kappa^2 r_0^2} \left( \frac{r_0}{r} \right)^3 \left\{1 +  \frac{9\phi_0}{5}\left(\frac{r_0}{r}\right)^2 \left[1- \frac{10}{9} \left(\frac{r_0}{r}\right) \right] \right\} \,, \nonumber
\end{eqnarray}
which tend to zero as $r \rightarrow \infty$. Note that assuming $\phi_0>0$, the energy density is positive throughout the spacetime, and at the throat takes the value $\rho(r_0)=3\phi_0/(5\kappa^2r_0^2)$. The NEC at the throat takes the form $\rho(r_0)+p_r(r_0)= (\phi_0-2)/(2\kappa^2 r_0^2)$, which is positive, by imposing $\phi_0>2$; it is an easy matter to verify that this specific example satisfies Eq. (\ref{NECeffSET2}). 

\section{Conclusions}\label{sect5}

%
The traditional approach of solving the Einstein field equation (EFE) consists in taking into account a plausible stress-energy tensor and, consequently deducing the geometrical structure of the spacetime. However, one can run the EFE in the reverse direction by first constructing the spacetime metric and then deduce the stress-energy tensor components. Wormhole physics is a specific example of adopting the latter reverse philosophy of solving the EFE. It was found that traversable wormholes possess a stress-energy tensor that violates the null energy condition at the throat. However, one may argue that this approach in finding the stress-energy components lacks a physical justification and motivation for the energy-momentum distribution. Furthermore, in the case of modified theories of gravity, in principle, one may show that it is the effective stress-energy tensor, containing higher order curvature terms, that supports these geometries. Thus, in modified gravity one may consider a specific meaningful equation of state for the normal matter threading the wormhole and that satisfies the energy conditions. In this manner, with the physically meaningful distribution of matter-energy at hand, one may in principle run the gravitational field equations in the traditional manner and solve for the remaining unknown functions. However, it is extremely difficult to provide exact solutions in the hybrid metric-Palatini theory outlined in the present paper, although it is possible to find a plethora of numerical solutions. Work along these lines is currently underway.

In concluding, our primary concern in this paper was to consider the possibility that wormholes be supported by the recently proposed hybrid metric-Palatini gravitational theory. We presented the general conditions for a wormhole relative to the null energy conditions at the throat and found specific solutions. In the first solution, we specified the redshift function, the scalar field and chose the potential that simplifies the modified Klein-Gordon equation. This solution is not asymptotically flat and needs to be matched to a vacuum solution. We emphasized that the surface stresses of the junction surface may, in principle, be imposed to satisfy the energy conditions. In the second example, we chose a particular scalar field and specified the metric functions, thus obtaining an asymptotically flat spacetime. We stress that the energy conditions for standard matter, in principle, may be imposed to hold while the structure and stability of wormhole solutions are guaranteed by the effective stress-energy induced by curvature terms. In fact, the issue of stability is extremely important and despite the fact that this was not a prime motivation for the present paper, the stability analysis is certainly a fundamental issue that will be explored in a future publication.

\section*{Acknowledgments}
SC is supported by INFN (iniziativa specifica NA12). TSK is supported by the Research Council of Norway.  FSNL acknowledges financial support of the Funda\c{c}\~{a}o para a Ci\^{e}ncia e Tecnologia through the grants CERN/FP/123615/2011 and CERN/FP/123618/2011. GO is supported by the Spanish grant FIS2008-06078-C03-02, FIS2011-29813-C02-02, the Consolider Program CPAN (CSD2007-00042), and the JAE-doc program of the Spanish Research Council (CSIC).

\end{document}